\documentclass{emulateapj}
\usepackage{apjfonts}
\usepackage{epsfig}

\usepackage[latin1]{inputenc}
\usepackage{lscape}

\shorttitle{Proper Motions in NGC~1333}
\shortauthors{Carrasco-Gonz\'alez et al.}

\begin{document}

\title{Proper Motions of Thermal Radio Sources near HH~7-11 in the NGC~1333 Star Forming Region}

\author{Carlos~Carrasco-Gonz\'alez\altaffilmark{1}, Guillem~Anglada\altaffilmark{1}, 
Luis~F.~Rodr\'{\i}guez\altaffilmark{2}, Jos\'e~M.~Torrelles\altaffilmark{3}, Mayra Osorio\altaffilmark{1}}

\altaffiltext{1}{Instituto Astrof\'{\i}sica Andaluc\'{\i}a, CSIC, Camino
Bajo de Hu\'etor 50, E-18008 Granada, Spain; charly@iaa.es,
guillem@iaa.es, osorio@iaa.es}

\altaffiltext{2}{Centro de Radioastronom\'{\i}a y Astrof\'{\i}sica UNAM,
Apartado Postal 3-72 (Xangari), 58089 Morelia, Michoac\'an, M\'exico;
l.rodriguez@astrosmo.unam.mx}

\altaffiltext{3}{Instituto de Ciencias del Espacio (CSIC)-IEEC, Facultat 
de F\'{\i}sica, Universitat de Barcelona, C/ Mart\'{\i} i Franqu\`es, 1, 
E-08028 Barcelona, Spain; torrelles@ieec.fcr.es}

\received{2008 June 4}
\accepted{2008 August 30}

\begin{abstract}

 Star forming regions are expected to show linear proper motions due to
the relative motion of the Sun with respect to the region. These proper
motions appear superposed to the proper motions expected in features
associated with mass ejection from the young stellar objects embedded in
them. Therefore, it is necessary to have a good knowledge of the proper
motions of the region as a whole in order to correctly interpret the
motions associated with mass ejection. In this paper we present the first
direct measurement of proper motions of the NGC~1333 star forming region.
This region harbors one of the most studied Herbig-Haro systems, HH 7-11,
whose exciting source remains unclear. Using VLA A configuration data at
3.6 cm taken over 10 years, we have been able to measure the absolute
proper motions of four thermal sources embedded in NGC~1333. From our
results we have derived the mean proper motions of the NGC~1333 star
forming region to be $\mu_{\alpha}\cos\delta$ = 9 $\pm$ 1 mas
yr$^{-1}$ and $\mu_{\delta}$ = $-$10 $\pm$ 2 mas~yr$^{-1}$. In this
paper, we also discuss the possible implications of our results in the
identification of the outflow exciting sources.

\end{abstract}


\section{Introduction}

 The stars in nearby regions of star formation are expected to show linear proper motions in the order of
milliarcseconds (mas) per year. These proper motions are the result of the relative motion between the Sun and
the stars studied, and can be obtained, to first order, by comparing the observed position of the star with
respect to the reference frame of the remote quasars at different epochs.

 In the last few years, Very Large Array (VLA) observations taken with time baselines of up to two decades have
been used to measure these proper motions for regions like Taurus (Loinard et al. 2003), L1527 (Loinard et al.
2002), L1551 (Rodr\'{\i}guez et al. 2003), Ophiuchus (Curiel et al. 2003), and Orion (G\'omez et al. 2005).

 Even when the VLA observations lack sufficient angular resolution to detect more subtle motions (i.e.
geometric parallax), they can provide not only the average proper motion of the region but also reveal the
presence of stars moving with peculiar velocities with respect to the region, such as those found in the Orion
BN/KL region (Rodr\'{i}guez et al. 2005, G\'omez et al. 2005), that suggest a runaway nature for some of the
sources. Much more accurate astrometry can be achieved using Very Long Baseline Interferometry (VLBI)
techniques (e.g. Loinard et al. 2007), but these observations are restricted to very compact and bright non
thermal radio stars and cannot be applied to thermal sources, where the emission is relatively extended and
its brightness temperature is not expected to exceed $10^4$~K, far below the sensitivity of present day VLBI.

\begin{deluxetable*}{cccccrc}
\tabletypesize{\scriptsize}
\tablewidth{0pt}
\tablecaption{Observations Parameters \label{tabla1}}
\startdata
\hline \hline
        &             &         &   Bootstrapped		    &			 &				    &	                         \\
	&	      &  VLA    &   Flux Density of		    & \multicolumn{2}{c}{Synthesized Beam\tablenotemark{b}} &	                         \\ \cline{4-5}
	& Observation & Project & Phase Calibrator\tablenotemark{a} &	      HPBW	 &   \multicolumn{1}{c}{P.A.}	    & rms Noise\tablenotemark{b} \\
 Epoch  & Date        &  Code   & (Jy)  			    &  (arcsec) 	 & \multicolumn{1}{c}{(deg)}	    & ($\mu$Jy beam$^{-1}$)	 \\ \hline
1989.1  & 89-Jan-14   & AR202   &      1.343 $\pm$ 0.003	    & 0.29 $\times$ 0.24 &         62                       &    28                      \\
1994.3	& 94-Apr-23   & AR277   &      1.686 $\pm$ 0.006	    & 0.28 $\times$ 0.20 &      $-$68                       &    27                      \\
1996.9	& 96-Dec-22   & AR277   &      1.305 $\pm$ 0.009	    & 0.35 $\times$ 0.30 &      $-$88                       &    24                      \\
1998.2	& 98-Mar-27   & AA218   &      1.476 $\pm$ 0.009	    & 0.26 $\times$ 0.24 &       $-$5                       &    18                      \\
1998.4	& 98-May-26   & AA218   &      1.610 $\pm$ 0.010	    & 0.26 $\times$ 0.24 &       $-$7                       &    18                      \\
1999.5  & 99-Jul-03   & AA239   &      1.681 $\pm$ 0.008	    & 0.29 $\times$ 0.25 &         39                       &    17                      \\
\hline
\enddata

\tablenotetext{a}{The phase calibrator used in all the epochs was 0333+321.}

\tablenotetext{b}{From naturally weighted maps.}

\end{deluxetable*}

\begin{deluxetable*}{crrcrr}
\tabletypesize{\scriptsize}
\tablewidth{0pt}
\tablecaption{Positions of the Sources\tablenotemark{a} \label{tabla2}}
\startdata
\hline \hline
          &                       \multicolumn{2}{c}{VLA~2}                           & &                   \multicolumn{2}{c}{VLA~3}                               \\ \cline{2-3} \cline{5-6}
Epoch     & \multicolumn{1}{c}{$\alpha$(J2000)} & \multicolumn{1}{c}{$\delta$(J2000)} & & \multicolumn{1}{c}{$\alpha$(J2000)} & \multicolumn{1}{c}{$\delta$(J2000)} \\ \hline
1989.1    & 03 29 01.9535 $\pm$ 0.0003      & 31 15 38.307 $\pm$ 0.005  	   & & 03 29 03.369 $\pm$ 0.002 	 & 31 16 01.89 $\pm$ 0.02	    \\
1994.3    &       01.9586 $\pm$ 0.0004      &       38.221 $\pm$ 0.004  	   & &       03.375 $\pm$ 0.002 	 &	 01.82 $\pm$ 0.02	    \\
1996.9    &       01.9605 $\pm$ 0.0004      &       38.285 $\pm$ 0.008  	   & &       03.376 $\pm$ 0.001 	 &	 01.75 $\pm$ 0.02	    \\
1998.2    &       01.9606 $\pm$ 0.0002      &       38.188 $\pm$ 0.005  	   & &       03.376 $\pm$ 0.001 	 &	 01.72 $\pm$ 0.02	    \\
1998.4    &       01.9604 $\pm$ 0.0002      &       38.211 $\pm$ 0.006  	   & &       03.373 $\pm$ 0.002 	 &	 01.78 $\pm$ 0.02	    \\
1999.5    &       01.9622 $\pm$ 0.0002      &       38.183 $\pm$ 0.003  	   & &       03.373 $\pm$ 0.001 	 &	 01.78 $\pm$ 0.01	    \\
\hline \\
          &                       \multicolumn{2}{c}{VLA~4A}                          & &                       \multicolumn{2}{c}{VLA~4B}                          \\ \cline{2-3} \cline{5-6}
Epoch     & \multicolumn{1}{c}{$\alpha$(J2000)} & \multicolumn{1}{c}{$\delta$(J2000)} & & \multicolumn{1}{c}{$\alpha$(J2000)} & \multicolumn{1}{c}{$\delta$(J2000)} \\ \hline
1989.1    & 03 29 03.730 $\pm$ 0.002	    & 31 16 04.02 $\pm$ 0.02		  & & 03 29 03.751 $\pm$ 0.002         & 31 16 04.08 $\pm$ 0.02     \\
1994.3    &       03.730 $\pm$ 0.002	    &	    03.95 $\pm$ 0.02		  & &	    03.758 $\pm$ 0.004         &       04.01 $\pm$ 0.02     \\
1996.9    &       03.731 $\pm$ 0.002	    &	    03.94 $\pm$ 0.02		  & &	    03.757 $\pm$ 0.002         &       04.01 $\pm$ 0.02     \\
1998.2    & \multicolumn{1}{c}{$\cdots$\tablenotemark{b}} & \multicolumn{1}{c}{$\cdots$\tablenotemark{b}} & & \multicolumn{1}{c}{$\cdots$\tablenotemark{b}} & \multicolumn{1}{c}{$\cdots$\tablenotemark{b}} \\
1998.4    &       03.735 $\pm$ 0.002	    &	    04.00 $\pm$ 0.03		  & &	    03.760 $\pm$ 0.002         &       03.99 $\pm$ 0.02     \\
1999.5    &       03.734 $\pm$ 0.002	    &	    03.96 $\pm$ 0.02		  & &	    03.757 $\pm$ 0.001         &       03.91 $\pm$ 0.02     \\
\hline
\enddata

\tablenotetext{a}{Positions derived from Gaussian ellipsoid fits in the naturally weighted maps. Units of right
ascension are hours, minutes and seconds, and units of declination are degrees, arcminutes, and arcseconds.}

\tablenotetext{b}{The components of the binary system VLA~4 could not be resolved in this epoch.}

\end{deluxetable*}  

\begin{figure*}
\epsscale{0.98}
\plotone{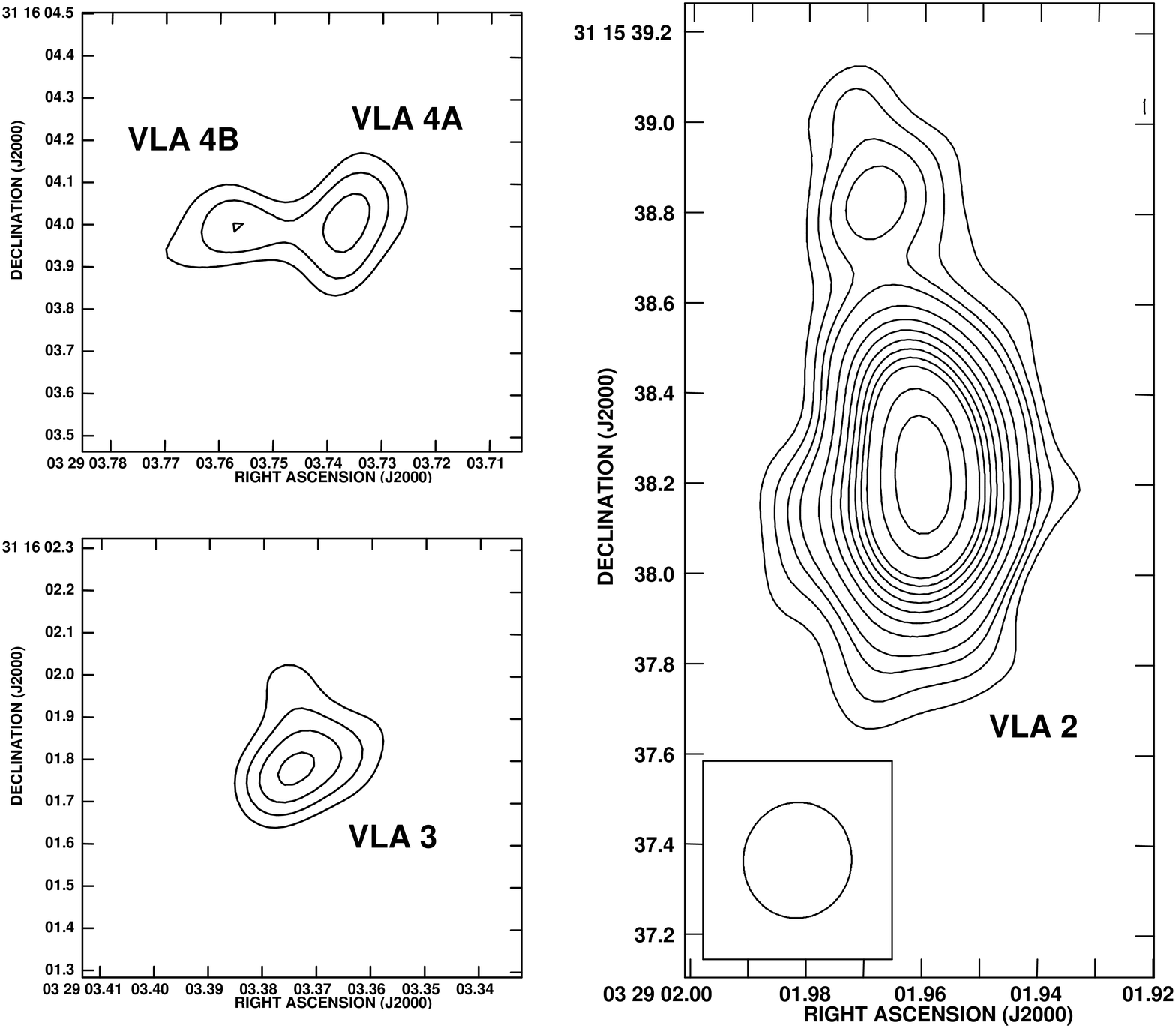}

\caption{\footnotesize{Naturally weighted VLA~3.6 cm continuum map of the thermal sources VLA~2, VLA~3, VLA~4A
and VLA~4B at the 1998.4 epoch. Contours are $-3$, 3, 4, 5, 6, 8, 10, 12, 14, 16, 18, 20, 25 and 30 times
the rms of the map, 18 $\mu$Jy beam$^{-1}$. The synthesized beam, shown in the right hand panel, is 0$\farcs$26
$\times$ 0$\farcs$24; P.A.=$-$7$^\circ$}. The three panels are plotted at the same scale.}

\label{fig1}
\end{figure*}

 The NGC~1333 star forming region, located at a distance of 235 $\pm$ 18 pc (Cernis 1990; Hirota et al. 2008)
in the Perseus complex, harbors the  classical bright Herbig-Haro (HH) system HH~7-11, first reported by Herbig 
(1974) and by Strom et al. (1974). The optically visible star SVS~13, discovered as a near-IR source (Strom et
al. 1976), is roughly aligned  with the chain of HH objects and was proposed as the powering source of  this HH
system. This association was questioned by Rodr\'{\i}guez et al. (1997), who discovered a cm radio source (VLA
3) located $6''$ to the SW  of SVS~13, and argued that this new object is a better candidate to drive the HH
outflow. SVS~13 is also associated with cm emission (source VLA~4  of Rodr\'{\i}guez et al. 1997) and mm
emission (Looney et al. 2000). Bachiller et al. (2000), through interferometric observations, found that SVS~13
is associated with an ``extremely'' high velocity molecular outflow, although ``standard'' velocity gas was
found in the vicinity of both SVS~13  and VLA~3. Subarcsecond VLA observations at 3.6 cm and 7 mm by Anglada et 
al. (2000, 2004) revealed that the radio source associated with SVS~13 is  actually a close binary with two
components, VLA~4A and 4B, separated by  0$\farcs$3 ($\sim$65 AU). Interestingly, a detailed analysis of the 
positions and spectral energy distribution in the cm-mm range suggests that the two sources have very different 
properties: VLA~4B appears to be associated with the observed strong mm emission,  probably arising from a
circumstellar dust disk, while VLA~4A is the  counterpart of the visible star SVS~13 with dust emission absent
or much  less significant in this component.

Since high velocity water masers are indicators of outflow activity and can be observed with high angular
resolution, Rodr\'{\i}guez et al. (2002) analyzed archive VLA data of the water masers associated with SVS~13 in
order to investigate which of the two stars of the binary system was the most likely candidate to drive the
outflow. It was found that the water masers appear segregated in two groups, according to their position and LSR
velocity. A group with the LSR velocity similar to that of the ambient cloud is associated with VLA~4A, and a
blueshifted velocity group is associated with VLA~4B. This result was interpreted as favoring VLA~4B as the
outflow driving source. However, the study of Rodr\'{\i}guez et al. (2002) was restricted to the line-of-sight
component of the velocity. Recently, Hirota et al. (2008) presented VLBI astrometric observations with VERA and
derived absolute positions and proper motions of the order of 15-20 mas~yr$^{-1}$ (corresponding to velocities
of 16-22 km~s$^{-1}$) of the water masers associated with SVS~13. These observations confirm the positional
association with VLA~4A of the masers with an LSR velocity close to that of the ambient cloud. However, the
proper motions (velocity and direction) relative to the sources could not be firmly established because of a
poor knowledge of the absolute proper motions of the region. In fact, the optical data of Herbig \& Jones (1983)
suggest that the stars projected upon heavy obscuration (presumably associated with the NGC 1333 cloud) have an
average proper motion of $\sim$10 mas yr$^{-1}$ relative to the stars near the edge of the obscuration
(presumably background stars). This suggests that the absolute proper motion of the region could have a
significant contribution to the proper motions of the masers observed by Hirota et al. (2008).

 In this paper we present an absolute astrometry analysis and proper motion calculations over 10 years of four
thermal radio sources in the NGC~1333 region. There are no reports of absolute proper motions for embedded young
stars in this region and our study provides the first determination of these motions.

\section{Observations}

 Data at 3.6 cm continuum were taken using the VLA of the National Radio Astronomy Observatory
(NRAO)\footnote{The NRAO is a facility of the National Science Foundation operated under cooperative agreement
by Associated Universities, Inc.}\ . The data consist of six epochs of observations, ranging from 1989 January
14 to 1999 July 3, obtained using the A configuration of the VLA, which provides an angular resolution of
$\sim$0$\farcs$3. A requirement to obtain reliable astrometry is that the same phase calibrator has been used
for most or preferably all observations. This requirement is accomplished in our case since all observations
were made using 0333+321 as phase calibrator. Data editing and calibration were carried out using the
Astronomical Image Processing System (AIPS) package of NRAO, following the standard VLA procedures. The
observation dates, the bootstrapped flux densities of 0333+321, as well as the parameters of the synthesized
beams, and the rms noise of the naturally weighted maps are given in Table \ref{tabla1}. Since the nominal
position of the phase calibrator used in the first epoch was slightly different from that used in the other five
epochs (the positions of the phase calibrators are periodically refined at the VLA), in order to obtain more
accurate astrometry, we corrected its position using the task CLCOR of AIPS before calibration of the first
epoch. Positions of the four thermal sources (VLA~2, VLA~3, VLA~4A, and VLA~4B; see Rodr\'{\i}guez et al. 1997
and Anglada et al. 2000 for the nomenclature of the sources) were derived in each epoch by Gaussian ellipsoid
fits in the naturally weighted maps.

\begin{figure*}
\plotone{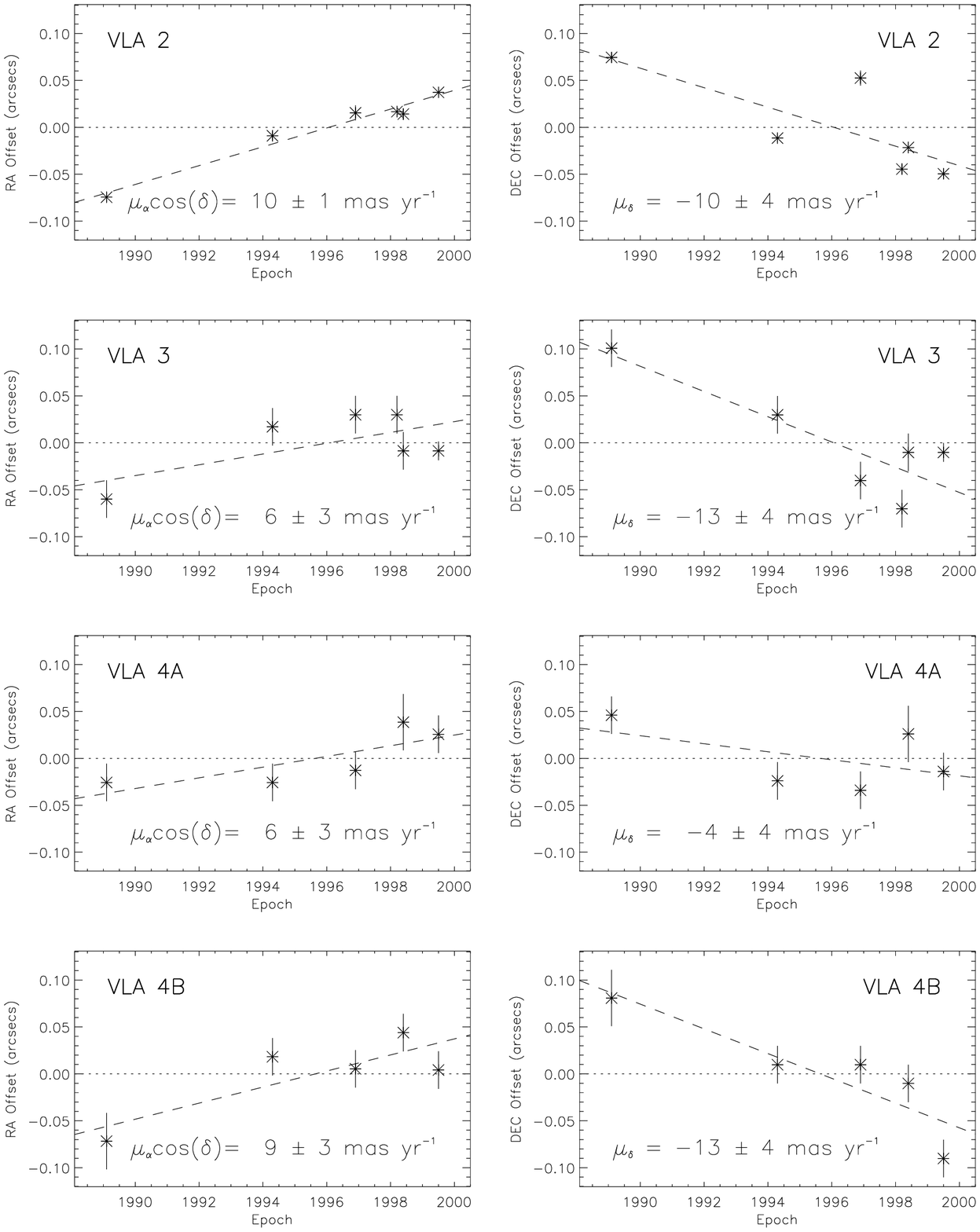}

\caption{\footnotesize{Position vs time diagrams of the sources VLA~2, VLA~3, VLA~4A, and VLA~4B. The positions
of each source are represented in the diagrams as right ascension and declination offsets (in arcsecs) relative
to the mean of the positions given in Table \ref{tabla2}. Error bars are obtained from Gaussian ellipsoid fits
in the naturally weighted maps. The actual errors could be greater because the morphology of the sources is not
exactly Gaussian (see text for details). The dashed line in each panel is a linear least squares fit to the
data. The values of $\mu_{\alpha}\cos\delta$ and $\mu_{\delta}$ obtained from these fits are also labeled in
each panel.}}

\label{fig2}
\end{figure*}
\begin{deluxetable*}{cccccc}
\tabletypesize{\scriptsize}
\tablewidth{0pt}
\tablecaption{Proper Motions of the sources \label{tabla3}}
\startdata
\hline \hline
	                        & $\mu_{\alpha}\cos\delta$  &     $\mu_{\delta}$    &   $\mu$           &      v           &  P.A.          \\
Source                          &   (mas~yr$^{-1}$)         &    (mas~yr$^{-1}$)    &  (mas~yr$^{-1}$)  &  (km~s$^{-1}$)   &  (deg)         \\ \hline
VLA~2                           &    10 $\pm$ 1             &    $-$10 $\pm$ 4      &   14 $\pm$ 3      &  16 $\pm$ 3      &  140 $\pm$ 10  \\
VLA~3                           &     6 $\pm$ 3             &    $-$13 $\pm$ 4      &   14 $\pm$ 4      &  16 $\pm$ 4      &  160 $\pm$ 10  \\
VLA~4A                          &     6 $\pm$ 3             &     $-$4 $\pm$ 4      &    7 $\pm$ 3      &   8 $\pm$ 3      &  120 $\pm$ 30  \\
VLA~4B                          &     9 $\pm$ 3             &    $-$13 $\pm$ 4      &   16 $\pm$ 4      &  18 $\pm$ 4      &  150 $\pm$ 10  \\ \hline
Weighted Mean\tablenotemark{a}  &    9 $\pm$ 1              &    $-$10 $\pm$ 2      &   14 $\pm$ 2      &  16 $\pm$ 2      &  140 $\pm$ 10  \\
\hline
\enddata

\tablenotetext{a}{The values of the total proper motion, velocity in the plane of the sky, and P.A. are derived from the mean values
of $\mu_{\alpha}\cos\delta$ and $\mu_{\delta}$.}

\end{deluxetable*}

\section{Results and Discussion}

 In Figure \ref{fig1} we show the 1998.4 maps of the sources VLA~2, VLA~3, VLA~4A, and VLA~4B. The measured
absolute positions of each source at each epoch are given in Table \ref{tabla2}. In Figure \ref{fig2} we present
position vs time diagrams for each source, and the proper motions, $\mu_{\alpha} \cos \delta$ and $\mu_{\delta}$,
obtained by linear least squares fits of these diagrams for each source are given in Table \ref{tabla3}. All the
sources show proper motions roughly to the SE with P.A. in the range from $\sim$120$^\circ$ to $\sim$160$^\circ$
and velocities in the range from 16 to 18 mas~yr$^{-1}$, except for VLA~4A, whose velocity is about one half the
velocity obtained for the other sources. This possible velocity difference of 8-10 km~s$^{-1}$ between VLA~4A and
the ambient cloud may account for the yet unexplained large increase in  the optical and near-IR brightness of
VLA~4A that took place in 1988-1990 (Eisl\"offel et al. 1991). The source may have simply displaced from a very
obscured region to a position where it became more easily detectable. However, because of the signal-to-noise
limitation of the data plus the errors produced by the intrinsic differences in the morphologies, orientations and
flux densities of the sources (see Fig. 1), we cannot conclude that the differences between the values obtained
for each source indicate statistically significant differences in the motions of the objects.

 As a representative value of the proper motion of the NGC~1333 region as a whole we adopt the weighted average
of the proper motion of the four sources, $\mu_{\alpha}\cos\delta$ = 9 $\pm$ 1 mas~yr$^{-1}$ and
$\mu_{\delta}$ = $-$10 $\pm$ 2 mas~yr$^{-1}$. Assuming a distance of 235~pc for NGC~1333, these values imply
a velocity of 16 $\pm$ 2 km~s$^{-1}$ along a P.A. of 140$^\circ$ $\pm$ 10$^\circ$. This is the first
measurement of the proper motion directly made on embedded objects that belong to the NGC~1333 star forming
region. Therefore, we believe that our value is, up to date, the most representative of the proper motion of NGC
1333.

 The proper motion of the NGC~1333 region obtained from our data can be compared with the value expected due to
the effect of the differential galactic rotation and the motion of the Sun with respect to the LSR. Assuming the
solar motion with respect to the LSR based on the Hipparcos satellite, (U$_0$, V$_0$, W$_0$)=(10.00, 5.25, 7.17)
km~s$^{-1}$ (Dehnen \& Binney 1998), the model of Brand \& Blitz (1993) for the galactic rotation, and a
distance of 235 pc for NGC~1333, we would expect proper motions of $\mu_{\alpha}\cos\delta$ = 2.9 mas~yr$^{-1}$
and $\mu_{\delta}$ = $-$7.5 mas~yr$^{-1}$, that imply a velocity of $\sim$9 km~s$^{-1}$ along a P.A. of
$\sim$160$^\circ$. This velocity is $\sim$7 km~s$^{-1}$ smaller than the value obtained from our data. Since the
difference between the proper motion predicted by the model and our measurements is of the order of the
one-dimensional rms velocity dispersion observed between molecular clouds (7.8 km~s$^{-1}$; Stark \& Brand
1989), we attribute this difference to the peculiar motion of the NGC~1333 cloud. We have looked for optical
stars associated with NGC~1333 with proper motions measured by Hipparcos. The only nearby star we find with
these characteristics is BD~+~30$^\circ$549, a B8:p type star that illuminates part of the reflection nebula
NGC~1333 (e.g. Rodr\'{\i}guez et al. 1990). For this star, Hipparcos reports proper motions of
$\mu_{\alpha}\cos\delta$ = 6.31$\pm$2.35 mas~yr$^{-1}$ and $\mu_{\delta}$ = -8.87$\pm$1.77 mas~yr$^{-1}$
(Perryman et  al. 1997). These values agree with the weighted average obtained by us at the $\pm$1-$\sigma$
level.

 Our results confirm that the source VLA~3 is a member of the NGC~1333 molecular cloud since this source shows
the same proper motion that the other members of the cloud. Although this conclusion alone does not resolve the
problem of the identification of the powering source of the HH 7-11 system, our results discard the possibility
that VLA~3 is a foreground or background source not related to the NGC~1333 region. Therefore, the proximity of
VLA~3 to the base of the HH 7-11 system is physically real, and not only a projection effect. 

 The determination of the proper motions of embedded young stars in NGC~1333 is useful to reanalyze the proper
motions of water masers reported by Hirota et al. (2008). These authors report VLBI observations of two maser
features associated with VLA~4A (SVS~13). They found absolute proper motions of the order of 15-20 mas~yr$^{-1}$
in a direction (P.A.$\simeq$114$^\circ$-133$^\circ$) roughly aligned with the jets and outflows in the region.
However, as noted by Hirota et al. (2008), these proper motions contain a significant contribution due to the
relative motion of the NGC~1333 region with respect to the Sun. In order to correct for the contribution due to
the relative motion of the NGC~1333 region with respect to the Sun, Hirota et al. (2008) subtracted the
theoretical proper motion of the region due to the Solar motion relative to the LSR (neglecting the contribution
of Galactic rotation, that they show is small). This correction results in proper motions of the order of 7-14
mas~yr$^{-1}$, and a significant change in the direction. The proper motions corrected in this way are not
parallel to the HH 7-11 jet and CO outflow. However, as we have discussed above, there is still a significant
contribution due to the peculiar proper motion of the NGC~1333 region that must be taken into account for a
proper interpretation of the results. 

 Our results show that the mean proper motion of the NGC~1333 region ($\mu_{\alpha}\cos\delta$ = 9$\pm$1 mas
yr$^{-1}$ and $\mu_{\delta}$ = $-$10$\pm$2 mas~yr$^{-1}$) determined from the four embedded thermal sources is
comparable to the absolute proper motions found by Hirota et al. (2008) for the two water maser features
associated with VLA~4A ($\mu_{\alpha}\cos\delta$ = 17.9$\pm$0.9 mas~yr$^{-1}$ and $\mu_{\delta}$ = $-$7.9$\pm$1.4
mas~yr$^{-1}$ for feature 1, and $\mu_{\alpha}\cos\delta$ = 10.6$\pm$1.7 mas~yr$^{-1}$ and $\mu_{\delta}$ =
$-$10.0$\pm$2.1 mas~yr$^{-1}$ for feature 2). Therefore, this correction appears to be very important to obtain
the true proper motions relative to the region. The close coincidence for feature 2 suggests that this maser is
almost stationary with respect to the cloud and that is not tracing significant peculiar proper motions. This
conclusion is supported by the fact that the LSR velocity of the masers (7-8 km~s$^{-1}$) nearly coincides with
the LSR velocity of the molecular cloud, as already noted by Rodr\'{\i}guez et al. (2002). After correction for
the proper motion of the NGC~1333 region determined by us, feature 1 has residual proper motions of
$\mu_{\alpha}\cos\delta$ = 9$\pm$2 mas~yr$^{-1}$ and $\mu_{\delta}$ = 2$\pm$3 mas~yr$^{-1}$. This indicate motions
to the east (P.A.$\simeq$90$^\circ$) with a velocity of $\sim$10 km~s$^{-1}$, that align poorly with the chain of
bright HH 7-11 objects and the molecular outflow. In any case, these small residual proper motions align much
better with the faint chain of HH objects (a, b, c and d) found by Davis et al. (1995), that aligns in the
west-east direction (see Fig. 2 of Rodr\'{i}guez et al. 1997 where the positions of the sources discussed in this
paper and the HH objects are also shown).

 We note, from the analysis discussed above, that we obtained proper motions of the maser features relative to
the NGC1333 cloud. An unambiguous measurement of the proper motions relative to VLA~4A would require to
subtract the individual proper motion of this source instead of the average motion of the NGC~1333 region. The
nominal proper motion velocity we obtained for this source is about one half the average value of the region.
Taken at face value, it could suggest that the VLA~4A source has a significant velocity relative to the ambient
cloud. However, the weakness of the source does not allow a reliable determination of its individual proper
motion with the currently available data. Additional, high sensitivity data would be required to investigate
possible motions of the VLA~4A source relative to the NGC~1333 cloud in order to accurately determine the
motions of the water maser features relative to VLA~4A.

\section{Conclusions}

 We have presented the results of multi-epoch high angular resolution VLA observations at 3.6 cm of the NGC~1333
star forming region. From these observations we have measured the absolute positions at different epochs of four
thermal sources (VLA~2, VLA~3, VLA~4A and VLA~4B) tracing YSOs embedded in this region. All these sources show
an absolute proper motion to the SE, which we interpret as a proper motion of the NGC~1333 region as a whole.
From our data we obtained the average values $\mu_{\alpha}\cos\delta$ = 9 $\pm$ 1 mas~yr$^{-1}$ and
$\mu_{\delta}$ = $-$10 $\pm$ 2 mas~yr$^{-1}$, that we interpret as representative of the proper motion of the
molecular cloud associated with the NGC~1333 region as a whole.

 Our results allow us to correct the proper motion measurements of water maser spots associated with VLA~4A (SVS
13) obtained by Hirota et al. (2008) in order to obtain the proper motions relative to the NGC~1333 cloud. We
conclude that feature 2 of Hirota et al. (2008) is practically stationary with respect to the cloud, and that
the small residual proper motions of feature 1 indicate motions to the east, suggesting that none of these
features is associated with the HH 7-11 system or the molecular outflow. 

 Our data also suggest that VLA~3 is a member of the NGC~1333 molecular cloud, since this source shows a proper
motion similar to that of the others sources embedded in the region. Since the position of VLA~3 appears close
to the base of the HH 7-11 system, we conclude that this proximity is physically real, and not a projection
effect. 

\emph{Acknowledgements.} C.C.-G. acknowledges support from MEC (Spain) FPU fellowship. G.A., C.C.-G., M.O., and
J.M.T. acknowledge support from MEC (Spain) grants AYA 2005-08523-C03 and AYA 2008-06189-C03 (including FEDER
funds), and from Junta de Andaluc\'{\i}a (Spain). L.F.R. acknowledges the support of DGAPA, UNAM, and of CONACyT
(M\'exico). We thank Tomoya Hirota for useful comments on this paper.

\newpage

\end{document}